# Unusual chemistry of the C–H–N–O system under pressure and implications for giant planets.

*Anastasia S. Naumova[1,2],\* Sergey V. Lepeshkin[1,2], Pavel V. Bushlanov[1] and Artem R. Oganov[1]*

[1] Skolkovo Institute of Science and Technology, Skolkovo Innovation Center, 3 Nobel St., Moscow, 143026, Russian Federation

[2] Lebedev Physical Institute, Russian Academy of Sciences, 119991 Leninskii prosp. 53, Moscow, Russia

\* naumova.nastasiya@gmail.com

**Abstract**

C–H–N–O system is central for organic chemistry and biochemistry, and plays a major role in planetary science (dominating the composition of "ice giants" Uranus and Neptune). The inexhaustible chemical diversity of this system at normal conditions explains it as the basis of all known life, but the chemistry of this system at high pressures and temperatures of planetary interiors is poorly known. Using *ab initio* evolutionary algorithm USPEX, we performed an extensive study of the phase diagram of the C–H–N–O system at pressures of 50, 200, and 400 GPa and temperatures up to 3000 K. Eight novel thermodynamically stable phases were predicted, including quaternary polymeric crystal $C_2H_2N_2O_2$ and several new N–O and H–N–O compounds. We describe the main patterns of changes in the chemistry of the C–H–N–O system under pressure and confirm that diamond should be formed at conditions of the middle-ice layers of Uranus and Neptune. We also provide the detailed $CH_4$–$NH_3$–$H_2O$ phase diagrams at high pressures, which are important for a further improvement of the models of ice giants – and point out that current models are clearly deficient. In particular, in existing models Uranus and Neptune are presented to have identical composition, nearly identical pressure-temperature profiles, and a single convecting middle layer ("mantle") made of a mixture $H_2O : CH_4 : NH_3 =$ 56.5 : 32.5 : 11. Here we provide new insights shedding light into the difference of heat flows from Uranus and Neptune, which require them to have different compositions, pressure–temperature conditions, and a more complex internal structure.

# 1. Introduction

The behavior of the C–H–N–O system is essential for processes in the interiors of giant planets such as Neptune and Uranus. Present-day models of these planets suggest a three-layer structure: an inner rocky core, a single massive middle "ice" layer, and outer H–He atmosphere. The middle layer is believed to contain water $H_2O$, methane $CH_4$, and ammonia $NH_3$[1] – but (except for the outermost layers [2]) not in the form of intact molecules, but rather products of their transformations. These transformation products are not well known. In the deep middle layer, the pressure and temperature are thought to vary from 20 to 600 GPa and from 2000 to 7000 K depending on the depth,[1] and chemistry is expected to be very different from the simple mixture $H_2O:CH_4:NH_3$. A long-standing puzzle is that Neptune (but not Uranus) radiates 2.61 times more energy than it receives from the Sun, and the origin of this excess heat is unknown.[3] One of the proposed explanations is the formation of diamond at high pressures, which is denser than other substances of the middle layer and gravitationally sinks inside the liquid planet and this sinking in strong gravitational field heats the planet (it is usually ignored, but light hydrogen, released in this process, may simultaneously go up, additionally releasing heat). This fascinating hypothesis was first proposed by Ross in 1981[4] and then supported experimentally: in laser-heated diamond anvil cells, methane decomposes to produce diamond and hydrogen at temperatures of 2,000–4,000 K and pressures up to 80 GPa.[5-7] In addition, during dynamic compression of polystyrene at high temperatures and ~150 GPa, carbon–hydrogen separation occurs with the formation of diamond.[8] Several theoretical studies of hydrocarbons[9-12] show that at high pressures methane becomes thermodynamically unstable and finally dissociates into diamond and hydrogen at ~300 GPa and zero temperature[9,10] or at > 300 GPa and above 4000 K,[12] which is consistent with the proposed hypothesis.

The described hypothesis of diamond formation in the interior of Neptune raises many questions. First, the process of diamond formation from hydrocarbons is driven both by pressure (due to high density of diamond) and temperature (due to high entropy of pure hydrogen). This means that the heat, produced as a result of diamond formation and sinking, will accelerate (rather than buffer) the process of diamond formation. Such self-accelerating process should have very quickly run itself to completion, very early in the history of Neptune. If so, all heat released by Neptune now is relict, produced shortly after the formation of the planet and still spent by it. Second, the same process should occur in Uranus (a planet thought to be very similar to Neptune in terms of size, mass, chemical composition, and pressure-temperature conditions) and one must expect similarly large excess of heat flow. However, Uranus irradiates only ~1.06 times more energy than it receives from the Sun. Thirdly, recent planetary studies give quite a different picture of Neptune's and Uranus' evolution and luminosity. There is an increasing evidence[13] that the currently used adiabatic models of ice giants cannot correctly describe the current luminosity of both Neptune and especially Uranus. According to the models, Neptune's and Uranus' current states could be achieved within the lifetime of the Solar system only if their initial temperatures were much lower than predicted by traditional models of their formation.[14]

The possible explanation of low heat flow on Uranus is the existence of a thermal boundary layer within its middle layer, separating very hot inner region from a much colder outer region. This would imply separate convection of these layers, with effectively only the outer region contributing to the heat flow – thus explaining Uranus' low luminosity and hotter interior.[15-17]

Yet another obscure issue is the unusual non-axisymmetric Uranian and Neptunian magnetic fields, which are assumed to be related to the unique internal structure. Magnetic fields of these planets are thought to originate from convection of their electrically conducting middle layers, the conductivity coming from ionic diffusion (at moderate pressures) or even electronic (metallic) conductivity in the very bottom of the middle layer.[18-20] It has been suggested that conductivity may arise from nearly complete ionization of water, the most abundant component of the layer.[21,22] Clearly, the chemistry of the middle layer is very different from just a mixture of $H_2O$, $CH_4$ and $NH_3$ molecules, and the ion carriers responsible for the unusual magnetic fields of Uranus and Neptune are yet to be found. Water and ammonia have been predicted to metallize deep inside the ice layer at 7000 and 5500 K, respectively, and 300 GPa,[20] which adds a contribution of the electronic conductivity to the generation of the magnetic field.

To shed light on these problems we performed an extensive study of the C–H–N–O system and stable compounds formed at high pressures. Due to very large number of possible stoichiometries, any 4-component system is extremely difficult for theoretical or experimental mapping of all stable compounds. There have been several studies of unary, binary, and ternary subsystems of C–H–N–O at high pressures, including the C–O,[23] C–H,[10,11] N–H,[24,25] C–N–O,[26] and H–N–O[27-32] systems, and works of our group on the phase diagrams of C–H,[9,10] C–H–O,[33] N–O,[34] C–N,[35,36] and O–H[37] systems. The full quaternary and some ternary systems have not been studied at high pressures. Besides, there are doubts about the completeness of the published data on the stable compounds due to complexity and structural diversity of such systems. In this work, we revise and summarize all known information and, taking advantage of the latest methodological developments in the USPEX code, perform an unprecedented, complex and computationally expensive, investigation of the thermodynamically stable phases in the quaternary C–H–N–O system and its ternary and binary subsystems at high pressures.

2. Methods

To predict thermodynamically stable compounds and structures, we used the evolutionary *ab initio* global optimization algorithm implemented in the USPEX code,[38,39] and performed calculations at pressures of 50, 200, and 400 GPa. To make predictions for the quaternary system more comprehensive, we started with three independent USPEX runs for each of six binary and four ternary subsystems at every target pressure. Initially each system was calculated from scratch (with 4–16 atoms for binary and 8–32 atoms in the primitive cell for ternary systems), and then we did runs for the same systems with the inclusion of all previously found structures as seeds. New structures were also produced by a recently developed random topological crystal structure generator.[40] After that, three USPEX runs for the quaternary C–H–N–O system were performed with 8–36 atoms per unit cell at each target pressure. The total number of structural relaxations in all USPEX calculations was ~1,800,000, which required significant supercomputer resources.

The total energy calculations and structure relaxations were performed using the PBE functional in the framework of the PAW method, with the plane wave kinetic energy cutoff of 850 eV and a uniform Γ-centered grid with $2\pi \times 0.056$ Å$^{-1}$ spacing for the reciprocal space sampling, using the VASP code.[41]

For the thermodynamically stable structures, we used the finite displacement method, as implemented in the PHONOPY code, to calculate the phonon dispersion curves, phonon density of states, and phonon contribution to the free energy. The dynamic stability of novel compounds was ascertained by the absence of imaginary frequencies in their phonon dispersion curves. Large ~10×10×10 Å supercells, hard potentials and an increased plane-wave kinetic energy cutoff of 1000 eV were adopted to avoid artificial imaginary frequencies. High-temperature calculations were done in the harmonic approximation, resulting in phase diagrams at each pressure and temperatures up to 3000 K, which can be found in Supporting Information, Section S2.

## 3. Results and Discussion

The resulting compositional phase diagrams of the C–H–N–O system at 50, 200, and 400 GPa and 0 K are shown in Figure 1. The phase diagrams without the zero-point energy correction and at 1000, 2000 and 3000 K were also calculated and are given in (Supporting Information, Section S1). A phase diagram of the quaternary system can be presented as a tetrahedron, with faces corresponding to ternary phase diagrams, and edges — to binary ones. Compared with previous studies, novel stable compounds were found in the N–O, H–N–O, and C–H–N–O systems. In total, we found 10 new structures, including eight thermodynamically stable ones and two compounds lying slightly above the convex hull (Figure 2). Below, we briefly describe the already known thermodynamically stable compounds and give a more detailed description of the new substances found in this work.

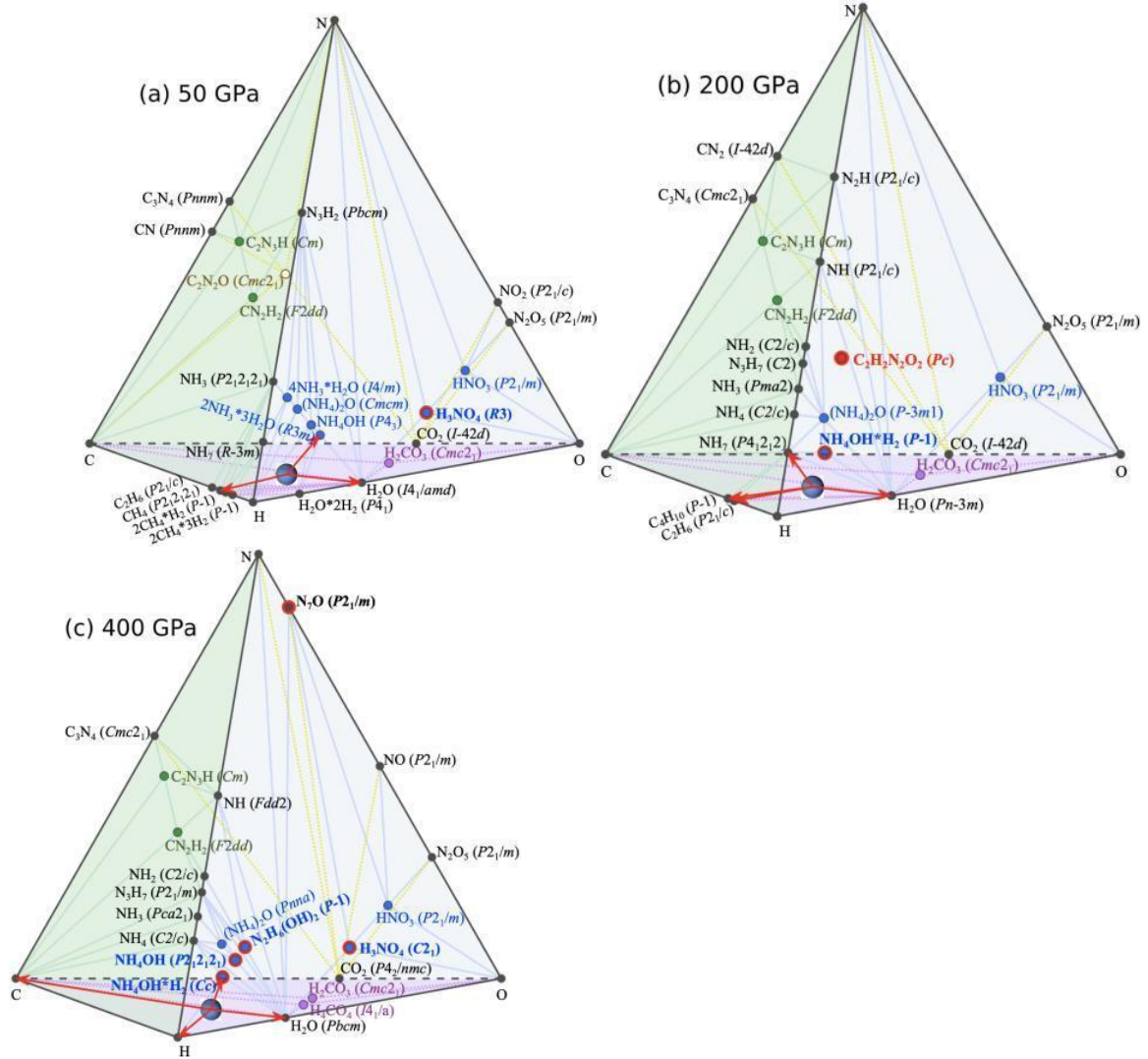

**Figure 1.** Phase diagrams of the C–H–N–O system at zero temperature and (a) 50, (b) 200, and (c) 400 GPa, taking into account the zero-point energy. The faces represent four ternary triangulated phase diagrams: stable H – N – O, C–H–N, C–H–O, and C–N–O compounds are marked by blue, green, violet, and yellow dots, respectively. Binary and pure compounds are shown by black dots. New stable compounds predicted in this work are marked by red circles. The currently accepted composition of Neptune's middle ice layer is shown by a large circle, with its components indicated by red arrows.

The C–O and H–O systems show quite simple chemistry: the only stable compounds are $CO_2$ ($I\bar{4}2d$ and $P4_2/nmc$) and $H_2O$ (ices VIII and X, and $Pbcm$ phase) at all target pressures and water/hydrogen cocrystal $H_2O \cdot 2H_2$ ($P4_1$) at 50 GPa.[23,37] C–N, C–H, N–H, and C–H–O systems have been exhaustively studied.[9,10,33,35,11,24,25,36,42] Our simulations reproduced all previously known compounds: CN (*Pnnm*), C₃N₄ (*Pnnm*, *Cmc*2₁, *Pnma*), CN₂ (*Ī42d*), CH₇ (*P̄1*), CH₅ (*P̄1*), CH₄ (*P2₁2₁2₁*), CH₃ (*P2₁/c*), C₂H₅ (*P̄1*), H₂CO₃ (*Cmc*2₁, *Cmc*2₁-II), H₄CO₄ (*I4₁/a*), NH (*C*2, *Fdd*2), N₂H (*P2₁/c*), N₂H₃ (*Pbcm*), NH₂ (*C*2/*c*), N₃H₇ (*C*2, *P*2₁/*m*), NH₃ (*P*2₁2₁2₁, *Pma*2, and *Pca*2₁), NH₄ (*C*2/*c*) and NH₇ (*R̄3m* and *P*4₁2₁2). In the C–N–O and C–H–N systems, only a few compounds are thermodynamically stable. *Cmc*2₁-C₂N₂O is known to exist at pressures from 20 to 100 GPa.[26] Two compounds, *F2dd*-CN₂H₂ and *Cm*-C₂N₃H, are shown to be stable throughout the studied pressure range. Despite the fact that these structures had been known before,[43] a comprehensive variable-composition search showing the absence of other thermodynamically stable compounds has not been performed until now.

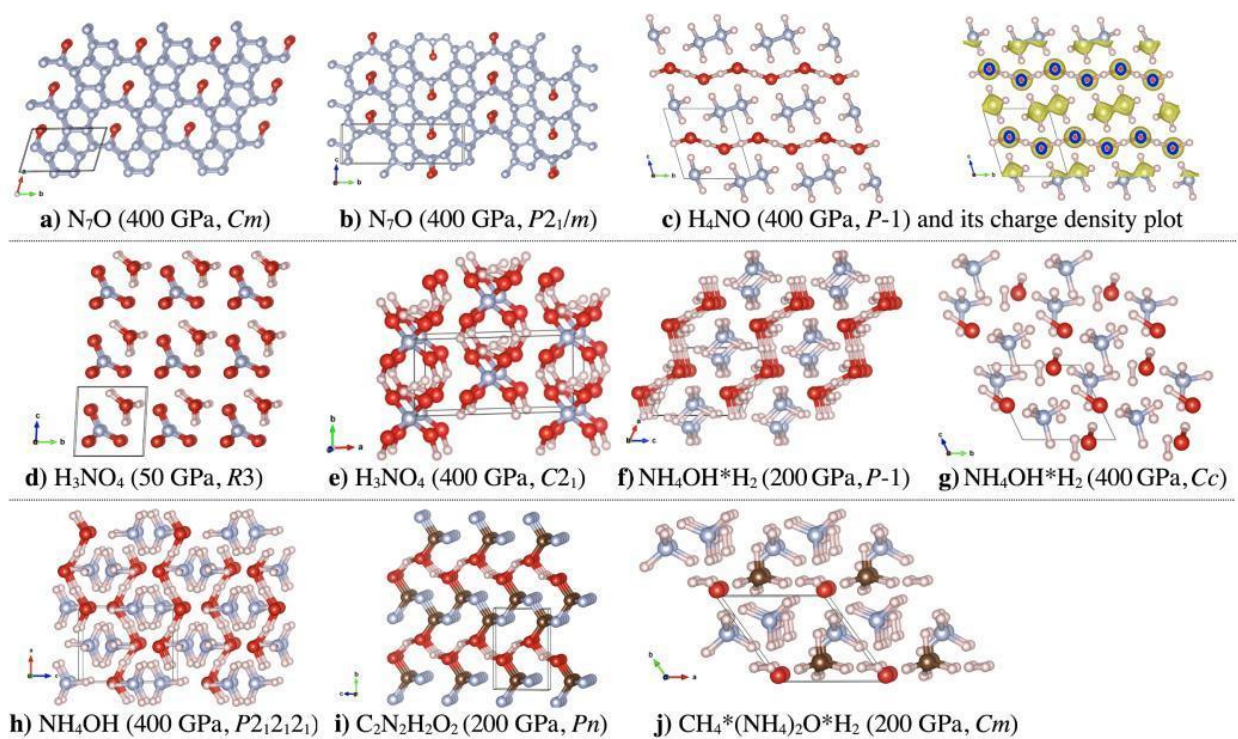

**Figure 2.** Predicted structures of the C–H–N–O phases. For H₄NO, the charge density plot is shown.

The N–O system has a rich chemistry of low-enthalpy metastable compounds, while only two compounds are thermodynamically stable. The calculations show that at 50 GPa, only *P*2₁/*m*-N₂O₅ and *P*2₁/*c*-NO₂ are stable, which is consistent with previously reported results;[34] at 200 and 400 GPa, NO₂ is unstable, but *P*2₁/*m*-NO and N₂O₅ are present on the phase diagram. A new composition, N₇O, was found to be thermodynamically stable at 400 GPa, adopting space group *Cm* or *P*2₁/*m* (Figure 2a,b) and consisting of 5- and 10-member N cycles that resemble a honeycomb structure with an oxygen atom inside the 10-member ones. These two structures are energetically very close, with an energy difference of about 3 meV/atom ($P2_1/m$ is slightly more preferable).

The H–N–O system, extremely diverse at high pressures, has been extensively studied in recent years (including works on $NH_3$–$H_2O$ mixtures),[27–32] and yet six new thermodynamically stable structures were discovered in our calculations: $P\bar{1}$-$H_4NO$, $R3$- and $C2_1$-$H_3NO_4$, $P\bar{1}$- and $Cc$-$NH_7O$ and $P2_12_12_1$-$NH_4OH$.

Nitric acid $P2_1/m$-$HNO_3$ is thermodynamically stable throughout the entire studied pressure range. $R3m$-$H_{12}N_2O_3$ and $I4/m$-$H_{14}N_4O$ are stable at 50 GPa, in agreement with literature.[27,29] Diammonium oxide $(NH_4)_2O$ has been predicted to adopt space group $Cmcm$ at 50 GPa, $P\bar{3}m1$ at 200 GPa, and $Pnna$ at 400 GPa,[27,29] which agrees with our calculations.

Several compositions from the previous study[27] become unstable according to our calculations. $C2/m$-$H_{10}N_2O$ and $R\bar{3}m$-$H_8N_4O$, which have been found to be thermodynamically stable at 50 GPa, and $P2_1/m$-$H_6N_2O$ (at 200 GPa) are not on the phase diagram anymore: our results showed that these phases lie about 6, 17, and 9 meV/atom above the convex hull, respectively.

The new unusual compound $P\bar{1}$-$H_4NO$ (Figure 2c) exists only at pressures about 400 GPa and is the salt of the positively charged dimer of ammonia and hydroxide anions, connected between each other into the infinite polymeric chain with symmetrical hydrogen bonding, which is seen in its charge density plot. This composition has been reported to adopt space group $C2/m$, forming a similar hydroxyl polymeric pattern, but our findings suggest the new $P\bar{1}$ phase to be ~28 meV/atom more stable.[28]

We found two new phases, $R3$ and $C2_1$, of $H_3NO_4$ (Figure 2d,e), which exist at 50 and 400 GPa, respectively. The calculations have shown this previously known compound to adopt space group $Pna2_1$ and be thermodynamically unstable at normal pressure.[44] At 50 GPa, this structure is ~10 meV/atom less stable than phase $R3$ discovered in this work.

Another cocrystal of ammonium hydroxide and a hydrogen molecule $NH_4OH•H_2$ appears on the phase diagram at about 200 GPa, adopts space group $P\bar{1}$ (Figure 2f), and contains a hydroxyl polymeric chain similar to that of $H_4NO$. After the transition into the nonpolymeric $Cc$ phase (Figure 2g), it remains stable up to at least 400 GPa.

Ammonium hydroxide $NH_4OH$ has been experimentally found to adopt several phases: $P2_12_12_1$ (0 GPa), $Pbca$ (0.5 GPa), $P4/nmm$ (12 GPa), $Pma2$ (50 GPa), and $Ima2$ (70 GPa).[30–32] In a recent theoretical work, this compound has been suggested to decompose at pressures above 60 GPa;[29] our predictions show that new high-pressure phase $P2_12_12_1$ (Figure 2h) is thermodynamically stable at 400 GPa.

In the C–H–N–O system calculations, we found one thermodynamically stable polymeric compound, $C_2H_2N_2O_2$, which adopts space group $Pn$ (Figure 2i) and is stable at 50 GPa and $T > 600$ K, while at 200 GPa it is stable in the entire studied temperature range, and at 400 GPa at $T > 1300$ K. Another discovered quaternary compound is a cocrystal of methane, molecular hydrogen, and ammonium oxide $CH_4•(NH_4)_2O•H_2$ (Figure 2j). At 50–200 GPa, it lies slightly above the convex hull by ~17 – 22 meV/atom, adopts space group $Cm$, and is dynamically stable (i.e. has no imaginary phonon frequencies). Several C–H–N–O molecular cocrystals lying slightly above the convex hull were also found in our calculations.

Summarizing the above, the C-H-N-O system is found to possess remarkable chemical diversity under pressure. Several patterns are important. Polymeric states are stabilized under pressure: water molecules and even hydroxyl ion polymerize with symmetric hydrogen bond formation: high-pressure forms of ice have structures with 3D bond connectivity, and hydroxyl ion under pressure is a linear chain! Methane $CH_4$ polymerizes too – forming longer and longer hydrocarbons on increasing pressure, and eventually diamond. Hydronitrogens tend to form ionic structures with $NH_4^+$, $N_2H_5^+$ and exotic $N_2H_7^+$ and $N_2H_6^{2+}$ cations and $NH_2^-$ anions, sometimes with symmetric hydrogen bonds as in $N_2H_{14}$ compound. Also, co-crystals which contain hydrogen molecule are likely to form (*e.g.* $NH_4OH•H_2$, $CH_4•(NH_4)_2O•H_2$).

Determination of the complete phase diagram of the C–H–N–O system at high pressure is of great interest for study of matter in the interiors of planets. Using the obtained phase diagrams, stable chemical compounds can be established for any given total composition. One of the recent models of Uranus' and Neptune's middle ice layer composition based on the Voyager mission data[45] postulates that the mass ratio of water, methane, and ammonia is 56.5:32.5:11. We found that at 50 GPa and zero temperature, Neptune's middle ice layer composition is a mixture of $P4_3$-$NH_4OH$, water, and methane; at 200 GPa — $R\bar{3}m$-$NH_3$, water, ethane and butane; at 400 GPa — the new compound $Cc$-$NH_4OH•H_2$, water, hydrogen, and crystalline diamond, which begins to form at pressures between 200 and 400 GPa. At temperatures of about 1000 K these mixtures do not change significantly. Although water, hydrogen, and the discovered H–N–O compounds become fluid at higher temperatures corresponding to the middle layer conditions of Neptune, this study of the well-defined crystalline matter gives important insights into their rich chemistry. Most probably, the predicted crystals melt, the hydrogen molecules of the found cocrystals evaporate into the atmosphere, the density of the middle ice layer increases (this is important for further refinements of models of Uranus and Neptune), and a molecular-ionic liquid is formed. At conditions of the middle layer this liquid will be a good ionic conductor containing at least the $OH^-$ and $NH_4^+$ ions (i.e. the environment is alkaline), which is consistent with previously reported observations.[20] If the real composition is different, also $H^+$, $NO_3^-$, $CO_3^{2-}$, $HCO_3^-$, $N_2H_6^{2+}$, $O^{2-}$ and several other ions could occur. Convection of such an electrically conducting liquid will generate a magnetic field.

Our study confirms M. Ross's suggestion about the diamond formation inside the ice giants and its sinking to the core, accompanied by the heat production.[4,46] However, as mentioned in the introduction, some important issues remain to be resolved - for example, the low luminosity of Uranus. At the same time, at least two different layers exist in the mantle of ice giants: outer layer containing a large concentration of hydrocarbons and inner layer strongly depleted in hydrocarbons. These layers will inevitably have very different properties (density, viscosity, thermal expansivity, etc.) and may convect separately – in which case there will be a thermal boundary layer across which heat is transferred only by (very inefficient) thermal conduction. This will reduce the amount of heat emitted by the planet and keep its deep interiors much hotter than one would expect from the adiabatic formula.

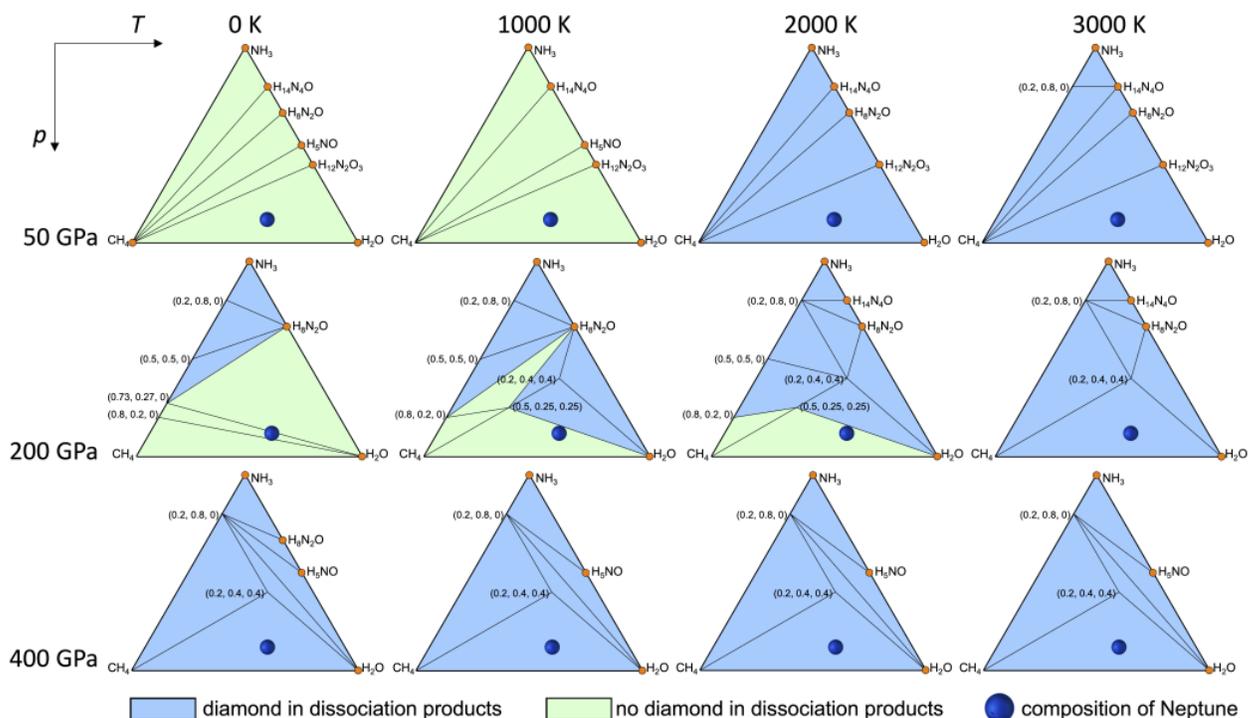

**Figure 3.** $CH_4$–$NH_3$–$H_2O$ slices of the quaternary C–H–N–O phase diagram at 50, 200, and 400 GPa and 0, 1000, 2000, and 3000 K.

In order to study the internal structure of the ice giants in more detail and explore the possible differences between Neptune and Uranus, we plot the $CH_4$–$NH_3$–$H_2O$ slices of the quaternary C-N-H-O phase diagrams. Figure 3 shows these diagrams at 50, 200, and 400 GPa and 0, 1000, 2000, and 3000 K; the diagrams at all temperatures with the step of 100 K with the dissociation products are given in Supporting Information, Section S3. The compositional areas containing diamond among the decomposition products are shown in blue, while diamond-free areas are colored green. In a nutshell, as pressure and temperature increase, diamond is more likely to be formed. At 50 GPa and 0 – 1700 K any mixture of methane, ammonia and water decomposes into diamond-free products, while at temperatures higher than 1700 K diamond is present in dissociation products also for any composition. At 400 GPa diamond is among dissociation products for any composition and temperature. The case of 200 GPa is the most interesting one: at 0 K the area of the phase diagram with diamond in its decomposition products occupies only a part of the entire diagram. As temperature rises, the area of this region increases and eventually occupies the entire diagram at T > 2000 K.

At 200 GPa and in the temperature range of ~1000-2000 K, the estimated composition of Uranus and Neptune lies very close to the boundary between diamond-bearing and diamond-free fields. If we assume only a small difference in the pressure–temperature conditions and/or composition of these planets, such that they fall one into a diamond-bearing and the other into diamond-free field, the planets would display quite different evolution. Note, however, that in very deep parts of the middle layer, both planets will again display the same processes. However, if these parts are shielded by a thermal barrier layer, heat produced there will be conserved in the deep interiors of Uranus and the observed luminosity will display no anomalies – as is the case. One should also keep in mind that although Neptune is slightly smaller than Uranus, it has a larger mass and a significantly (29 %) greater density than Uranus, and one might imagine that the amount of diamond (and heat) produced in Neptune is simply larger than that in Uranus, and it takes longer to spend it out. Greater density is also consistent with more hydrogen having escaped from the interiors as a product of diamond formation.

However, the real situation in the ice layers might be more complicated, because phase composition and pressure-temperature profile inside Uranus' and Neptune's ice layers are known very approximately. In particular, the temperature is quite high and the substance is most likely a liquid or is in so called superionic state, which is characterized by diffusive protons in solid lattices of heavy nuclei.[20,47] But hypotheses based on solid state calculations proposed in our study may remain valid.

## 4. Conclusions

We studied the C–H–N–O system in a wide pressure range (50–400 GPa) and its highly diverse chemistry. Among the novel compounds we found $Cm$- and $P2_1/m$-$N_7O$, cocrystal $Cc$-$NH_4OH \cdot H_2$, an exotic $NH_4O$ structure that can be described as $P\bar{1}$-$N_2H_6(OH)_2$, two new polymeric phases $R3$- and $C2_1$-$H_3NO_4$, and quaternary polymer $Pc$-$C_2H_2N_2O_2$. All these structures are both thermodynamically and dynamically stable. We also found a peculiar metastable cocrystal $Cm$-$CH_4 \cdot (NH_4)_2O \cdot H_2$. Knowledge of the thermodynamically stable C–H–N–O compounds at high pressures allowed the construction of its phase diagram for the first time. $CH_4$ – $NH_3$ – $H_2O$ slices are of particular interest for improving the models of middle ice layers of Neptune and Uranus, since they are assumed to consist of methane, ammonia and water. Our results confirm the hypothesis of diamond formation and its role in producing excessive heat in Neptune and shed light into the hitherto mysterious differences in heat production on Uranus and Neptune and their unusual magnetic fields.

**Acknowledgement**

Structure predictions were performed with support from Russian Science Foundation (grant 19-72-30043), phase diagrams were investigated with support from Russian Ministry of Science and Higher Education (grant 2711.2020.2 to leading scientific schools) and Russian Foundation for Basic Research (grant 19-02-00394).

**References**

1. Hubbard, W. B. & MacFarlane, J. J. Structure and evolution of Uranus and Neptune. *J. Geophys. Res.* **85**, 225–234 (1980).
2. Guillot, T. Interiors of giant planets inside and outside the solar system. *Science* **286**, 72–77 (1999).
3. Pearl, J. C. & Conrath, B. J. The albedo, effective temperature, and energy balance of Neptune, as determined from Voyager data. *Journal of Geophysical Research* vol. 96 18921 (1991).
4. Ross, M. The ice layer in Uranus and Neptune—diamonds in the sky? *Nature* **292**, 435–436 (1981).
5. Hirai, H., Konagai, K., Kawamura, T., Yamamoto, Y. & Yagi, T. Polymerization and diamond formation from melting methane and their implications in ice layer of giant planets. *Phys. Earth Planet. Inter.* **174**, 242–246 (2009).
6. Benedetti, L. R. *et al*. Dissociation of CH4 at high pressures and temperatures: diamond formation in giant planet interiors? *Science* **286**, 100–102 (1999).
7. Zerr, A., Serghiou, G., Boehler, R. & Ross, M. Decomposition of alkanes at high pressures and temperatures. *High Press. Res.* **26**, 23–32 (2006).
8. Kraus, D. *et al*. Formation of diamonds in laser-compressed hydrocarbons at planetary interior conditions. *Nature Astronomy* **1**, 606–611 (2017).
9. Gao, G. *et al*. Dissociation of methane under high pressure. *J. Chem. Phys.* **133**, 144508 (2010).
10. Naumova, A. S., Lepeshkin, S. V. & Oganov, A. R. Hydrocarbons under Pressure: Phase Diagrams and Surprising New Compounds in the C–H System. *J. Phys. Chem. C* **123**, 20497–20501 (2019).
11. Liu, H., Naumov, I. I. & Hemley, R. J. Dense Hydrocarbon Structures at Megabar Pressures. *J. Phys. Chem. Lett.* **7**, 4218–4222 (2016).
12. Ancilotto, F. Dissociation of Methane into Hydrocarbons at Extreme (Planetary) Pressure and Temperature. *Science* vol. 275 1288–1290 (1997).
13. Podolak, M., Helled, R. & Schubert, G. Effect of non-adiabatic thermal profiles on the inferred compositions of Uranus and Neptune. *Mon. Not. R. Astron. Soc.* **487**, 2653–2664 (2019).
14. Fortney, J. J. & Nettelmann, N. The Interior Structure, Composition, and Evolution ofáGiant

Planets. *Space Sci. Rev.* **152**, 423–447 (2010).

15. Nettelmann, N. *et al*. Uranus evolution models with simple thermal boundary layers. *Icarus* **275**, 107–116 (2016).
16. Leconte, J. & Chabrier, G. Layered convection as the origin of Saturn's luminosity anomaly. *Nat. Geosci.* **6**, 347–350 (2013).
17. Vazan, A., Helled, R., Podolak, M. & Kovetz, A. The evolution and internal structure of Jupiter and Saturn with compositional gradients. *ApJ* **829**, 118 (2016).
18. Chau, R., Hamel, S. & Nellis, W. J. Chemical processes in the deep interior of Uranus. *Nat. Commun.* **2**, 203 (2011).
19. Redmer, R., Mattsson, T. R., Nettelmann, N. & French, M. The phase diagram of water and the magnetic fields of Uranus and Neptune. *Icarus* **211**, 798–803 (2011).
20. Cavazzoni, C. *et al*. Superionic and metallic states of water and ammonia at giant planet conditions. *Science* **283**, 44–46 (1999).
21. Kirk, R. L. & Stevenson, D. J. Hydromagnetic constraints on deep zonal flow in the giant planets. *Astrophys. J.* **316**, 836–846 (1987).
22. Nellis, W. J. *et al*. The nature of the interior of Uranus based on studies of planetary ices at high dynamic pressure. *Science* **240**, 779–781 (1988).
23. Iota, V. *et al*. Six-fold coordinated carbon dioxide VI. *Nat. Mater.* **6**, 34–38 (2007).
24. Steele, B. A. & Oleynik, I. I. Pentazole and Ammonium Pentazolate: Crystalline Hydro-Nitrogens at High Pressure. *J. Phys. Chem. A* **121**, 1808–1813 (2017).
25. Qian, G.-R. *et al*. Diverse Chemistry of Stable Hydronitrogens, and Implications for Planetary and Materials Sciences. *Sci. Rep.* **6**, 25947 (2016).
26. Steele, B. A. & Oleynik, I. I. Ternary Inorganic Compounds Containing Carbon, Nitrogen, and Oxygen at High Pressures. *Inorg. Chem.* **56**, 13321–13328 (2017).
27. Steele, B. A. Computational Discovery of Energetic Polynitrogen Compounds at High Pressure. (University of South Florida, 2018).
28. Shi, J., Cui, W., Botti, S. & Marques, M. A. L. Nitrogen-hydrogen-oxygen ternary phase diagram: New phases at high pressure from structural prediction. *Physical Review Materials* vol. 2 (2018).
29. Robinson, V. N., Marqués, M., Wang, Y., Ma, Y. & Hermann, A. Novel phases in ammonia-water mixtures under pressure. *The Journal of Chemical Physics* vol. 149 234501 (2018).
30. Loveday, J. S. & Nelmes, R. J. The ammonia hydrates—model mixed-hydrogen-bonded systems. *High Press. Res.* **24**, 45–55 (2004).
31. Mafety, A. Etude ab initio des glaces d'ammoniac fluoré et hydraté sous conditions thermodynamiques extrêmes. (Université Pierre et Marie Curie-Paris VI, 2016).
32. Fortes, A. D. *et al*. Crystal Structure of Ammonia Monohydrate Phase II. *Journal of the American Chemical Society* vol. 131 13508–13515 (2009).
33. Saleh, G. & Oganov, A. R. Novel stable compounds in the CHO ternary system at high pressure. *Sci. Rep.* **6**, 1–9 (2016).
34. Li, D. *et al*. Nitrogen oxides under pressure: stability, ionization, polymerization, and superconductivity. *Sci. Rep.* **5**, 16311 (2015).
35. Dong, H., Oganov, A. R., Zhu, Q. & Qian, G.-R. The phase diagram and hardness of carbon nitrides. *Sci. Rep.* **5**, 9870 (2015).
36. Pickard, C. J., Salamat, A., Bojdys, M. J., Needs, R. J. & McMillan, P. F. Carbon nitride frameworks and dense crystalline polymorphs. *Phys. Rev. B Condens. Matter* **94**, 094104 (2016).

37. Qian, G.-R., Lyakhov, A. O., Zhu, Q., Oganov, A. R. & Dong, X. Novel Hydrogen Hydrate Structures under Pressure. *Scientific Reports* vol. 4 (2015).
38. Oganov, A. R. & Glass, C. W. Crystal structure prediction using ab initio evolutionary techniques: principles and applications. *J. Chem. Phys.* **124**, 244704 (2006).
39. Oganov, A. R., Lyakhov, A. O. & Valle, M. How Evolutionary Crystal Structure Prediction Works  and Why. *Acc. Chem. Res.* (2011).
40. Bushlanov, P. V., Blatov, V. A. & Oganov, A. R. Topology-based crystal structure generator. *Comput. Phys. Commun.* **236**, 1–7 (2019).
41. Kresse, G. & Furthmüller, J. Efficiency of ab-initio total energy calculations for metals and semiconductors using a plane-wave basis set. *Comput. Mater. Sci.* **6**, 15–50 (1996).
42. Song, X. *et al*. Exotic Hydrogen Bonding in Compressed Ammonia Hydrides. *J. Phys. Chem. Lett.* **10**, 2761–2766 (2019).
43. Salamat, A. *et al*. Tetrahedrally bonded dense $C_2N_3H$ with a defective wurtzite structure: X-ray diffraction and Raman scattering results at high pressure and ambient conditions. *Physical Review B* vol. 80 (2009).
44. Huck, P., Jain, A., Gunter, D., Winston, D. & Persson, K. A Community Contribution Framework for Sharing Materials Data with Materials Project. *2015 IEEE 11th International Conference on e-Science* (2015) doi:10.1109/escience.2015.75.
45. Hubbard, W. B. & Marley, M. S. Optimized Jupiter, Saturn, and Uranus interior models. *Icarus* vol. 78 102–118 (1989).
46. Kraus, D. On Neptune, It's Raining Diamonds. *Am. Sci.* **106**, 285–288 (2018).
47. Robinson, V. N. & Hermann, A. Plastic and superionic phases in ammonia–water mixtures at high pressures and temperatures. *J. Phys. Condens. Matter* **32**, 184004 (2020).